\documentclass[10pt,preprint]{article}

\usepackage{newtxtext,newtxmath}

\usepackage{graphicx}

\usepackage[letterpaper,left=0.5in, right=0.5in, top=1in, bottom=1in]{geometry}

\usepackage{xcolor}

\renewenvironment{abstract}
	{\quotation}
	{\endquotation}

\date{}

\makeatletter
\renewcommand{\fnum@figure}{\textbf{Figure \thefigure}}
\renewcommand{\fnum@table}{\textbf{Table \thetable}}
\makeatother

\usepackage{scicite}

\usepackage{url}

\usepackage{braket}

\def\scititle{
Cavity-mediated hybridization of several molecules\\ in the strong coupling regime
}

\title{\bfseries \boldmath \scititle}

\author{
	Jahangir~Nobakht$^{1,2}$,
	André~Pscherer$^{1,2}$,
	Jan~Renger$^{1}$,
	Stephan~G{\"o}tzinger$^{1,2,3}$,
	Vahid~Sandoghdar$^{1,2\ast}$ \\
	\small$^{1}$Max Planck Institute for the Science of Light, D-91058, Erlangen, Germany.\\
	\small$^{2}$Department of Physics, Friedrich-Alexander University, D-91058, Erlangen, Germany.\\
	\small$^{3}$Graduate School in Advanced Optical Technologies (SAOT), Friedrich-Alexander University, D-91052, Erlangen, Germany.\\
	\small$^\ast$Corresponding author. Email: vahid.sandoghdar@mpl.mpg.de
} 
\begin{document} 


\maketitle

\begin{abstract} \bfseries \boldmath
Molecular complexes are held together via a variety of bonds, but they all share the common feature that their individual entities are in contact. In this work, we introduce and demonstrate the concept of a \textit{molecular optical bond}, resulting from the far-field electromagnetic coupling of several molecules via a shared mode of an optical microcavity. We discuss a collective enhancement of the vacuum Rabi splitting and study super- and sub-radiant states that arise from the cavity-mediated coupling both in the resonant and dispersive regimes. Moreover, we demonstrate a two-photon transition that emerges between the ground and excited states of the new optical compound. Our experimental data are in excellent agreement with the predictions of the Tavis-Cummings Hamiltonian and open the door to the realization of hybrid light-matter materials.
\end{abstract}


\noindent
Hybridization of molecular bonds commonly takes place when the electronic clouds of the different entities overlap \cite{herman1983recent}. Electrostatic dipole-dipole coupling has also been known to hybridize molecular energy levels at separations much smaller than a wavelength of light \cite{Kasha1963}. This class of interactions has intrigued scientists for its central role in natural systems such as light harvesting complexes and pigment aggregates \cite{jang2018delocalized, Hestand2017, Son2021} as well as technological applications involving energy transfer and singlet fission \cite{Kasha1963, Scholes2003, Smith2013}. Over the last decades, a variety of natural and synthetic platforms have been employed to explore and exploit such near-field collective effects \cite{Hettich2002, zhang2016visualizing, Boulais2018}, but far-field hybridization of molecules at much larger distances has not been reported previously. In this work, we demonstrate that the energy levels of isolated molecules can be hybridized via a common mode of an optical microcavity. The resulting coupling of the individual molecules to each other invokes the concept of \textit{molecular optical bonds}.

When a two-level emitter is placed in a cavity, its radiative properties are modified due to the change in the density of states at its transition frequency $\omega$. Within the language of cavity quantum electrodynamics (CQED) \cite{haroche2006exploring}, the coupling can be described by a cooperativity parameter $C=4g^2/\kappa\gamma$, where $\kappa$ and $\gamma$ represent the energy decay rates (linewidths) of the cavity and the emitter, respectively. Here, $g=\mu\sqrt{\omega /2\epsilon \hbar V}$ denotes the coupling strength for a cavity of mode volume $V$ and emitter transition dipole moment $\mu$ embedded in a medium of dielectric function $\epsilon$. In the  \textit{weak coupling regime}, the emitter radiation rate is enhanced by $C$-fold, but the photon quickly leaks out of the cavity. In the \textit{strong coupling regime}, where $4g>\kappa+\gamma$, energy is coherently exchanged between the emitter and the cavity mode \cite{raizen1989normal, khitrova2006vacuum}. These phenomena have been demonstrated both at the ensemble and single-emitter levels in a variety of contexts over the last three decades \cite{haroche2006exploring,Flick2017,Hirai2023}. A new challenge is to study and resolve the states that emerge from the coupling of a controlled number of emitters to a cavity, e.g., in the context of quantum state engineering and polaritonic chemistry \cite{Becher2023, Flick2017, Hirai2023}. 

\section{Experimental platform}\label{sec2}

\begin{figure}[!htb]
  \centering
  \includegraphics[width=0.4\textwidth]{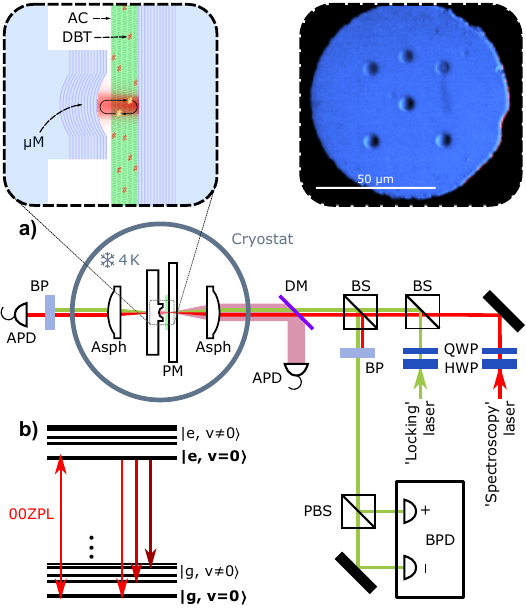}
  \caption{\textbf{Experimental scheme.} \textbf{a)} A continuous-wave (CW) spectroscopy laser beam is coupled into the cryogenic microcavity. The transmitted laser light is detected with a photon-counting avalanche photodiode (APD). On the reflection side, we also have the possibility of detecting the red-shifted fluorescence photons with a second APD. During the measurement, a CW locking laser beam at $\lambda$ = 700 nm is used to monitor cavity length changes. Inset left: Close-up view of the microcavity arrangement: a DBT-doped anthracene crystal is placed between a flat and a curved mirror at the end of a pedestal. Inset right: front view of a pedestal with six microfabricated curved mirrors; one is chosen for experimentation. Asph: aspheric lens, DM: dichroic mirror, BP: bandpass filter, BS: non-polarizing beamsplitter, PBS: polarizing beamsplitter, $\mu$M: micromirror, PM: planar mirror, QWP: quarter-wave plate, HWP: half-wave plate, BPD: balanced photodiode. \textbf{b)} Schematics of the energy levels of a DBT molecule, indicating the vibrational manifolds of the ground and excited states.}
  \label{fig:setup}
\end{figure}

Figure\,\ref{fig:setup}a) depicts the schematics of the experimental setup, the basic features of which were discussed in our previous publications \cite{Wang2019,Pscherer2021}. The center piece of the experiment consists of a plano-concave Fabry-Perot microcavity that contains an anthracene (AC) crystal of submicrometer thickness extended over a lateral dimension of $\sim 200$ $\mu$m (see inset in Fig.\,\ref{fig:setup}a)). The AC crystal is doped with dibenzoterrylene (DBT) molecules, which belong to the family of polycyclic aromatic hydrocarbons (PAH) and have previously been featured in many quantum and nano-optical studies \cite{Hettich2002, moerner1999illuminating, Hwang2009, Wang2019, Toninelli2021, Pscherer2021}. DBT possesses a strong zero-phonon line (00ZPL) that connects the vibrational ground levels of the electronic ground and excited states (see Fig.\,\ref{fig:setup}b)). When doped in AC, the 00ZPL of DBT occurs at a wavelength of $\lambda\sim784$\,nm, has a Fourier-limited linewidth of $ \gamma/2\pi \sim 40$\,MHz at $T\lesssim4$\,K, and carries $\simeq \frac{1}{3}$ of the total emission out of the excited state (denoted by branching ratio, $\alpha$), thus, exhibiting a high degree of coherence \cite{Nicolet2007, Wang2019}. As in other solid-state systems, the transition frequencies of the individual DBT molecules embedded in AC are spread over an inhomogeneous bandwidth, which was about 400\,GHz in our case. 

To improve the mechanical stability of the cryogenic setup over our previous experiments, we exploited a slight angle between the flat cavity mirror and the pedestal carrying the micromirror (see inset in Fig.\,\ref{fig:setup}a)) to bring the two in gentle contact. This eliminated the need for actively locking the cavity frequency, while allowing for sufficient axial motion through actuation of a piezoelectric element to tune the cavity resonance within tens of GHz \cite{flaagan2022diamond}. Nevertheless, we used the Hänsch-Couillaud error signal obtained from the locking laser (see Fig.\,\ref{fig:setup}a)) to monitor the cavity stability \cite{Pscherer2021}. The beam from a narrow-band continuous-wave Ti:Sapph laser (spectroscopy beam) was coupled to the microcavity from the flat mirror side, and its transmission through the microcavity was recorded as the laser frequency was scanned. The data reported in this paper were collected from two cavities with different band edge frequencies for the flat distributed Bragg reflector (DBR). For the majority of the data, the red-shifted fluorescence was cut out. A full width at half-maximum (FWHM) linewidth of $\kappa/2\pi=3.5$\,GHz on the $8^{\rm th}$ longitudinal cavity mode yielded a finesse of $\sim$13,500. For the two-photon measurements, the flat DBR was designed to be partially transparent for the red-shifted fluorescence emission of DBT. In our previous works, we reported a substantial Purcell factor that led to the strong modification of the branching ratio in a single molecule \cite{Wang2019}, demonstrated the onset of single-molecule strong coupling, and presented several nonlinear optical effects at the single photon level \cite{Pscherer2021}. Such single-molecule studies are usually performed at low DBT:AC doping to allow spectral isolation of single molecules within the observation volume and the inhomogeneous band of DBT. In this work, we increased the DBT doping level to arrive at the regime, where the 00ZPLs of several molecules coincided with a microcavity resonance.




\section{Two molecules coupled to a single cavity mode}
\begin{figure*}[!htb]
  \centering
  \includegraphics[width=0.7\textwidth]{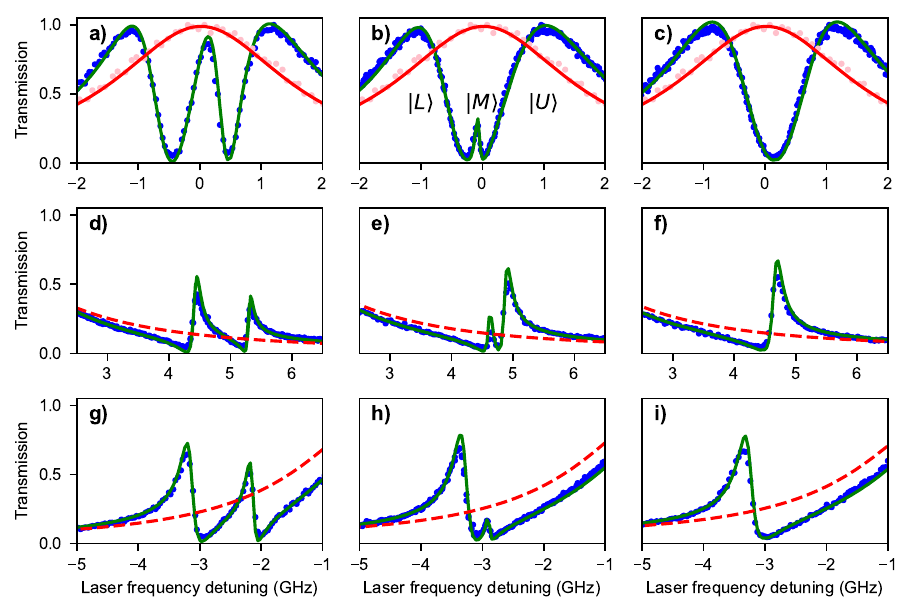}
  \caption{\textbf{Cavity transmission when coupled to two molecules.} Blue symbols show experimental data while the green curves represent the results of theoretical fits. The pink symbols and the red curve illustrate the measured spectrum of the bare cavity and a Lorentzian fit with a linewidth of \( 3.46 \)\,GHz, respectively. Coupling strengths were $ g_1 / 2\pi = 0.82 \pm 0.01$\,GHz and $g_2 / 2\pi = 0.6 \pm 0.02$\,GHz. Three distinct detuning values between the molecules are presented: $\delta_{12} /2\pi= 0.91 \pm 0.01$\,GHz \textbf{a)}, $0.24 \pm 0.01$\,GHz \textbf{b)}, and $0.05 \pm 0.04$\, GHz \textbf{c)}. \textbf{d-f)} and \textbf{g-i)}  show the cavity tuned to the red and blue sides, respectively. In each case, the cavity resonance is set at the origin of the horizontal axis. Red dashed curves display the extrapolated empty cavity transmission spectrum.}
  \label{fig:3config_res}
\end{figure*}

We start by modeling the interaction of two molecules with a single mode of a cavity. This is achieved by extending the Jaynes-Cummings Hamiltonian to the Tavis-Cummings Hamiltonian \cite{TavisCummings1968},
\begin{align}\label{JC}
\begin{split}
 \frac{\mathcal{H}}{\hbar} &=  \omega_1  {\sigma_1}^\dagger \sigma_1 + \omega_2  {\sigma_2}^\dagger \sigma_2 +  \omega_c a^{\dagger} a\\
 &+  g_1 (a {\sigma_1}^\dagger + a^{\dagger} \sigma_1 ) + g_2 (a {\sigma_2}^\dagger + a^{\dagger} \sigma_2 ),
\end{split} 
\end{align}
where $\sigma_i$ ($\sigma^\dagger_i$) ($i\in\{1,2\}$) and $a$ ($a^\dagger$) are the lowering (raising) operators for the two emitters and the field, respectively. The parameter $\omega_c$ represents the cavity frequency, $\omega_i$ is the transition frequency of the $i$-th molecule, and $g_i$ describes its coupling strength to the cavity mode. In the simplified scenario of a cavity that is resonant with two identical molecules with ground and excited states $\ket{g_i}$ and $\ket{e_i}$, we obtain
\begin{subequations}
\begin{align}
\ket{L}&= (\ket{g_1, e_2}+\ket{e_1, g_2})\ket{0}-\ket{g_1, g_2} \ket{1}, \\ 
\ket{M}&= (\ket{g_1, e_2}-\ket{e_1, g_2}) \ket{0}, \\
\ket{U}&= (\ket{g_1, e_2}+\ket{e_1, g_2})\ket{0}+\ket{g_1, g_2} \ket{1}
\end{align}
\end{subequations}
as new (unnormalized) polaritonic modes, where $\ket{0}$ and $\ket{1}$ represent the cavity field with zero and one photon, respectively. Each state can be described as a coherent superposition of three excitations. For the upper polariton state ($\ket{U}$), all excitations are in phase, whereas for the lower polariton state ($\ket{L}$), the molecular excitations are in phase with each other but out of phase with the photonic excitation. In the case of the middle polariton state ($\ket{M}$), the molecular excitations are out of phase with each other, canceling the overall material dipole moment, resulting in a dark state that does not couple to the cavity mode.

The red curve in Fig.\,\ref{fig:3config_res}a) displays the fit to the cavity transmission spectrum measured far from the resonances of any molecules in the sample. The blue symbols show the transmission spectrum when the curved mirror was moved to tune $\omega_c$ to the region of $\omega_1$ and $\omega_2$, spaced by $\frac{\delta_{1,2}}{2\pi}= \frac{\omega_1-\omega_2}{2\pi}=0.91 \pm 0.01$\,GHz. Here, $\delta_{1,2}$ is large enough that the two molecules can be considered to be independent of each other. Various cases presented in Fig.\,\ref{fig:3config_res}b-i) examine the coupling for different cases of $\delta_{1,2}$ and $\Delta_i=\omega_i-\omega_c$. 

\subsection{The resonance regime}
Oscillators generally couple most efficiently when they share their resonance frequencies, i.e., when $\Delta_{1}=\Delta_{2}=0$. To achieve this resonance condition, $\omega_{1,2}$ can be ideally dialed in a controlled fashion, e.g., by effectuating a local Stark effect on the DBT molecules \cite{rattenbacher2023chip}. Given the absence of microelectrodes in our sample, however, we chose an alternative approach, which consisted of illuminating a molecule by a strong laser beam with an effective power of $\sim 1$\,mW in the cavity, equivalent to an intensity $I\simeq 6\times10^4\,\rm{W/cm^2}$. This resulted in a frequency shift of that molecule's 00ZPL. We attribute this effect to a slight local modification of the AC crystal in the neighborhood of the molecule caused by a small amount of energy that is deposited in it, similar to a phenomenon reported in recent publications \cite{Colautti2020, lange2024superradiant}. Figure\,\ref{fig:3config_res}b) shows an example, where this approach was used to reduce $\delta_{1,2}/2\pi$ to 240\,MHz. The two peaks on the left and right represent transitions to $\ket{U}$ and $\ket{L}$, and the small sharp feature in the middle reports on the subradiant state $\ket{M}$ that is not fully dark.

In Fig.\,\ref{fig:3config_res}c), $\delta_{1,2}$ was further reduced to reach the resonance condition within 50\,MHz. On resonance, the new hybridized eigenfrequencies read $\omega_L=\omega_c -\sqrt{{g_1}^2+{g_2}^2}$, $\omega_M=\omega_c$, and $\omega_U=\omega_c + \sqrt{{g_1}^2+{g_2}^2}$, respectively. Here, one only observes two peaks for transitions to $\ket{U}$ and $\ket{L}$ because the dark state $\ket{M}$ does not couple to the cavity mode. In the simplest case of two identical molecules with decay rates $\gamma _0$ and coupling strengths $g_0$, the effective coupling of two molecules becomes $\sqrt 2 g_0$, and we obtain $\gamma _L = \gamma _U= (\gamma_0+\kappa)/2$. Although state $\ket{M}$ does not couple to the cavity mode, its coupling to the remaining free-space modes yields an overall decay rate $\gamma _M=\gamma_0$.


The full dynamics of the density operator is governed by a master equation. We obtained theoretical fits (solid curves) to the measured data (symbols) presented in Fig.\,\ref{fig:3config_res} from a numerical simulation that incorporates the coherent unitary evolution of the Tavis-Cummings Hamiltonian, the decay of the cavity mode, and the decay of molecules into free space. As $g_1$ and $g_2$ depend on $V$ and on the position of the molecules within the cavity mode, they do not change if $\omega_c$ is varied by a small amount. This allows us to extract $g_1$ and $g_2$ from fitting the data in Fig.\,\ref{fig:3config_res}a). Next, we use this information to fit the spectra in Fig.\,\ref{fig:3config_res}b,c) and to deduce $\omega_1, \omega_2$. The fits let us extract $\kappa/ 2\pi=3.46 \pm 0.03$\,GHz, $g_1 / 2\pi = 0.82 \pm 0.01$\,GHz and $g_2/ 2\pi=0.60 \pm 0.02$\,GHz. In this case, $g_1$ and $g_2$ lie just below the exceptional point of the Jaynes-Cummings Hamiltonian at $\frac{1}{4}(\kappa + \gamma_0) / 2\pi = 0.87$\,GHz \cite{Choi2010}, but the collective enhancement of the vacuum Rabi splitting given by $\sqrt{g_1^2 + g_2^2}/ 2\pi=1.02$\,GHz (see Fig.\,\ref{fig:3config_res}c)) enters the regime of two-molecule strong coupling. 

\subsection{The dispersive regime}
In the dispersive regime, where $|\Delta_i| \gg \kappa$, molecules cannot dissipate their energies into the cavity mode because the resonance condition is not met. Figure\,\ref{fig:3config_res}d-f) illustrates that if we tune the frequencies of two initially decoupled bare molecules (panel d)), we first observe the formation of a partially sub-radiant and a bright super-radiant state (panel e)). Eventually, when the two molecules are resonant with each other, the sub-radiant state becomes fully dark (panel f)). In other words, the two molecules hybridize via the cavity mode although they are not resonant with it. We elaborate on this in the next section. The dashed lines in the figure display the tail of the unperturbed cavity resonance, and the solid green curves present excellent theoretical fits to the experimental data. As illustrated in Fig.\,\ref{fig:3config_res}g-i), the sign of the coupling reverses if the cavity resonance is tuned to the blue side of the molecular pair.

\section{Cavity-mediated molecule-molecule coupling}
The composite arrangement of two molecules in a high-finesse cavity not only strongly affects the transmission of an external laser beam, it also presents a setting in which the two molecules influence each other via the shared cavity mode. Assuming weak excitation in the dispersive regime, where the cavity is not occupied by photons, Eq.\,(\ref{JC}) simplifies to \cite{Majer2007,Blais2021}
\begin{align}\label{Eq:dispersiveH}
\begin{split}
\frac{\mathcal{H}}{\hbar} &= \Tilde{\omega}_1 {\sigma_1} ^{\dagger} \sigma_1 + \Tilde{\omega}_2 {\sigma_2} ^{\dagger} \sigma_2 + J_{12} ({\sigma_1} ^{\dagger} \sigma_2 + \sigma_1 {\sigma_2} ^{\dagger}).
\end{split} 
\end{align}
The new frequencies $\Tilde{\omega}_i = \omega_i + \frac{2 g_i^2}{\Delta_i}$ and $\Tilde{\omega}_c=\omega_c-(\frac{2 g_1^2}{\Delta_1}+\frac{2 g_2^2}{\Delta_2})$ express the influence of the molecules and the cavity on each other. The coupling parameter $J_{12} = \frac{g_1 g_2}{\Delta_1}+\frac{g_1 g_2}{\Delta_2}$ represents the interaction energy of the two molecules through exchange of virtual photons via the common cavity mode. In analogy with the emitter-cavity interaction, one can also invoke the concept of cooperativity given by $C_{12}=\frac{4J_{12}^2}{\gamma_1 \gamma_2}$ to describe the efficiency of the interaction between the two molecules. When including a factor of $\alpha^2$ to account for the branching ratios of the molecules, we obtain $C_{12}=13,\,30$ for the cases in Figs.\,\ref{fig:3config_res}f) and \ref{fig:3config_res}i), respectively. 

The observed sub- and super-radiant phenomena in Fig.\,\ref{fig:3config_res}d-i) closely resemble the scenario of two molecules that undergo dipole-dipole coupling in the near field \cite{Hettich2002, Trebbia2022, lange2024superradiant}. The sign of the detuning $\Delta$ determines the energy ordering of the super- and subradiant states and corresponds to that of J ($\Delta>0$) and H ($\Delta<0$) aggregates, respectively \cite{lange2024superradiant, Trebbia2022}. Theoretical fits depicted by the solid curves in Figs.\,\ref{fig:3config_res}f) and \,\ref{fig:3config_res}i) let us extract $J_{12}/ 2\pi = 220 \pm 10$ and $-330 \pm 10$\,MHz, respectively. This is comparable with the dipole-dipole coupling rate of two resonant molecules separated by 25\,nm (H configuration) and 28\,nm (J configuration), respectively. 

\begin{figure}[!htb]
  \centering
  \includegraphics[width=0.35\textwidth]{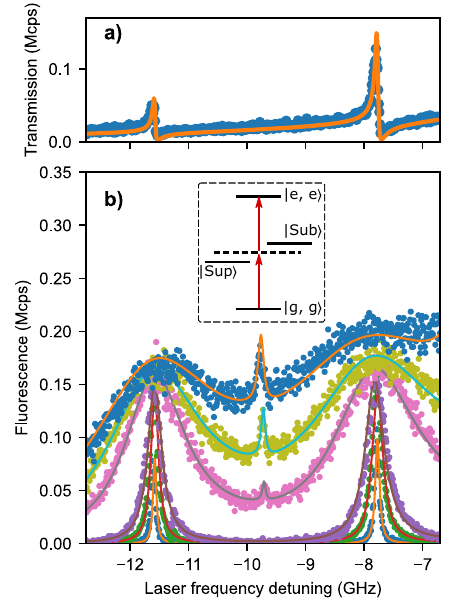}
  \caption{\textbf{Two-photon transition.} \textbf{a)} Off-resonance transmission spectrum of the cavity in the presence of two molecules that are dispersively coupled to it. The vertical axis is expressed in the units of mega counts per second (Mcps). The x-axis is shared with b). The origin marks the cavity frequency. \textbf{b)}  A series of fluorescence spectra recorded from the cavity at increasing laser powers. At higher laser powers, a two-photon peak appears in the middle of the sub- and super-radiant states. The excitation powers correspond to $0.2$, $2$, $10$, $199$, $474$, $950$\,photons per cavity life time, in the increasing order, respectively. Inset shows the energy diagram of the coupled system and the two-photon transition (red arrows).}
  \label{twophoton}
\end{figure}

Previous studies have shown that an interesting consequence of the hybridization between two molecules in the near field is the emergence of a two-photon transition to the fully excited state $\ket{e_1,e_2}$. This transition takes place at higher powers and appears midway between the super- and sub-radiant states \cite{Hettich2002, Trebbia2022, lange2024superradiant}. To investigate such a two-photon transition in our system, we increased the laser power. Figure\,\ref{twophoton}a) shows the cavity transmission spectrum for two dispersively-coupled molecules with $\delta_{12}/2\pi= 3.8$\,GHz, $\kappa/2\pi= 1.58$\,GHz, $g_1/2\pi=0.7$\,GHz and $g_2/2\pi=0.72$\,GHz. The symbols in Fig.\,\ref{twophoton}b) display a series of spectra recorded from the same frequency interval by recording the fluorescence signal from the flat mirror side of the cavity. It is evident that as the laser power is increased, the original resonances become power broadened and a narrow peak appears midway. The high quality of the fits provided by the solid curves and the power-dependence of the new resonance confirm that the new transition is a property of the new optical compound that results from the hybridization of two discrete molecules.

\begin{figure*}[!htb]
  \centering
    \includegraphics[width=0.7\textwidth]{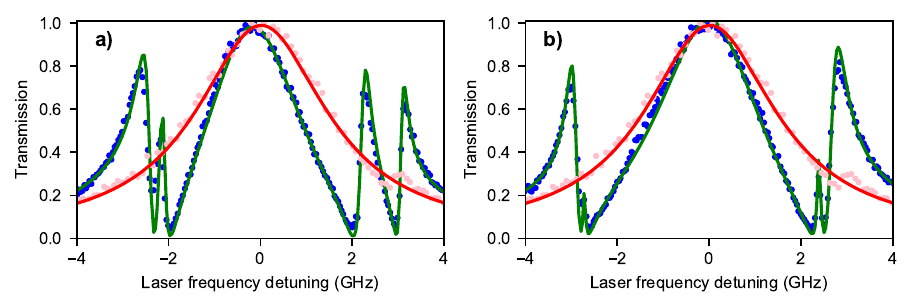}
      \includegraphics[width=0.7\textwidth]{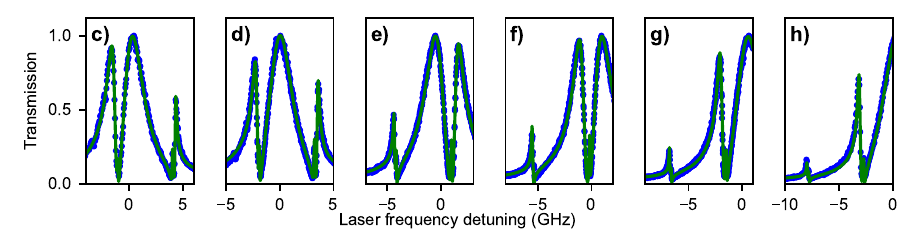}
    \includegraphics[width=0.7\textwidth]{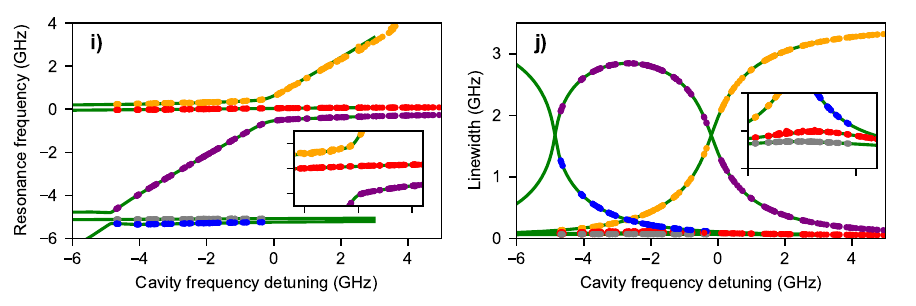}
   \includegraphics[width=0.7\textwidth]{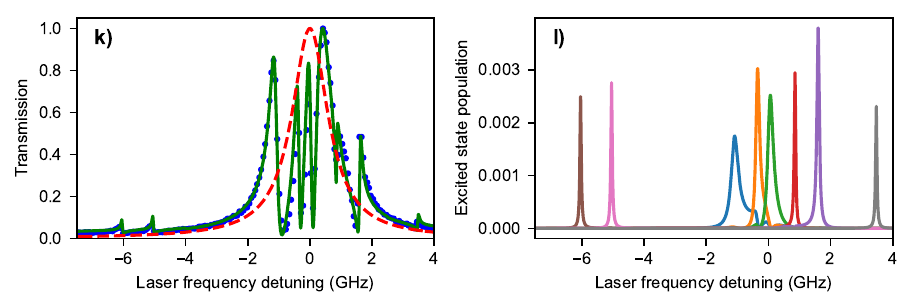}
  \caption{\textbf{Coupling many molecules.} \textbf{a)} Two molecular pairs coupled to the cavity mode. \textbf{b)} Same as a) but with smaller frequency differences within each pair. \textbf{c-h)} Snap shots of the coupled system in b) for different cavity frequency detunings. \textbf{i)} Evolution of the five resonance frequencies in c-h). Red and grey data points represent subradiant states. Inset shows a close-up of the region around the origin, set to the resonance of the cavity with the blue-detuned pair. \textbf{j)} Same as i) but for resonance linewidths. Inset shows a close-up of the region around -3\,GHz. \textbf{k)} Symbols show the cavity transmission spectrum resulting from the coupling of eight molecules. The dashed curve shows the bare cavity resonance for comparison, placed at the origin. The solid curve presents the theoretical fit, which allows us to extract $g_i /2\pi= 0.5, 0.25, 0.25, 0.1, 0.3$\,GHz for five of the molecules. \textbf{l)} Calculated excited-state populations of the eight molecules.}
  \label{fig:4molecule}
\end{figure*}

\section{Up to eight molecules}

A decisive advantage of our experimental system is the relative ease with which we can increase the number of molecules that couple via a common mode of the cavity. Two key experimental features enable this. First, molecules can be doped at a fairly high level, reaching average separations that are well below100\,nm. Second, the scannable micromirror of the cavity allows one to explore different sections of the thin AC crystal, thus facilitating the identification of favorable regions in the sample. Figure\,\ref{fig:4molecule}a) displays the cavity transmission after dispersive coupling to four molecules, whereby one pair lies on the blue and the other on the red side of $\omega_c$. The solid curves show the theoretical spectra fitted to the experimental data. The high fit quality lets us deduce $\Delta_1,\,\Delta_2,\,\Delta_3,\,\Delta_4=2\pi \times$ (2.18, 3.12, -2.16, -1.81)\,GHz, respectively. Figure\,\ref{fig:4molecule}b) displays the outcome when the frequency differences of the molecules within each pair are further reduced. 

It is instructive to investigate the resonance frequencies and linewidths of the various features in the spectrum of the coupled system of Fig.\,\ref{fig:4molecule}b) as the cavity frequency is scanned. Figure\,\ref{fig:4molecule}c-h) shows a few snap shots. In panel c), the cavity is near resonant with the two red-detuned molecules, and in panel f), it becomes resonant with the blue-detuned pair. Figures\,\ref{fig:4molecule}d,e) and \ref{fig:4molecule}g,h) show the cases where $\omega_c$ lies between the two pairs and when the cavity resonance is detuned to the blue side of both pairs, respectively. In Fig.\,\ref{fig:4molecule}i), we present the frequency evolution of the five resonances that result from the four molecules and the cavity as a function of $\omega_c$. Here, we observe an anticrossing (yellow and purple) as the cavity resonance traverses the blue-detuned pair, giving rise to the upper and lower polaritons. The central peak (red), which signifies the sub-radiant mode does not experience a notable change. However, we find a small frequency shift for the super-radiant states (purple and yellow, see inset), which we attribute to the collective Lamb shift induced by the cavity vacuum field. A similar but weaker phenomenon takes place as $\omega_c$ crosses the second pair on the red-detuned side. In Fig.\,\ref{fig:4molecule}j), we also plot the linewidths of the various resonances. The data clearly show the broadening of the molecular resonances as they hybridize with the cavity mode to form polaritonic states. The inset shows the close-up view of the slight linewidth change that occurs for the subradiant states.

The symbols in Fig.\,\ref{fig:4molecule}k) present an example, where eight molecules are coupled to the cavity mode in the resonant or near-resonant manner. Tuning the various molecular frequencies via laser illumination was not straightforward in this case, but future integration of microelectrodes on the flat cavity mirror should allow us to explore the complex parameter space of the Dicke states \cite{Dicke1954}. To gain more insight into the data, we computed the excited-state population of the individual molecules based on the outcome of the very good fit presented in Fig.\,\ref{fig:4molecule}k). The deviations from Lorentzian line shapes in Fig.\,\ref{fig:4molecule}l) indicate the onset of molecule-molecule coupling.  

\section{Discussion and outlook}\label{sec13}

We have presented the laboratory realization of a platform, where a discrete number of molecules located at distances much larger than their physical size experience an \textit{optical bond}. To achieve this, we coupled them to a common mode of a microcavity, entering the strong coupling regime. This new paradigm is made possible by the high doping density and spectral quality of PAHs in a thin crystal coupled to a scannable Fabry-Perot microcavity. Investigation of a well-defined number of interacting emitters has also been of great interest in quantum technology and quantum optics at large, but despite several important advances \cite{Majer2007, Mlynek2014, casabone2015enhanced, reimann2015cavity, kaufman2015entangling, Evans2018, Samutpraphoot2020, Lodahl2023}, scaling to large numbers has remained elusive. In this work, we showed that in addition to collective effects of the molecule-cavity system on an external laser beam, the energy levels of the individual molecules are hybridized, giving rise to sub- and super-radiant states. Furthermore, we demonstrated a two-photon transition that emerges when two molecules hybridize to form a super-molecule as an optical compound.

Several improvements can be implemented in future efforts to explore and advance far-field molecular interactions. First, one can decrease the radius of curvature of the micromirror and reduce the cavity length \cite{Kelkar2015} to increase the molecule-cavity coupling strength $g$ by about one order of magnitude. Other resonant architectures based on photonic crystals or plasmonic nanostructures promise to offer even higher coupling strengths \cite{Evans2018, Gonzalez-Tudela2011, Chikkaraddy2016, Matsuzaki2017, Liu2024}. Moreover, integration of microelectrodes and the use of nanocrystals \cite{pazzagli2018self,musavinezhad2024high} will offer more precise control of each molecular frequency and will facilitate addressing a larger number of molecules. Some of the desirable studies entail mapping the spectra of all the Dicke states that result from the coupling of $N$ molecules \cite{Dicke1954,hertzog2019strong} and exploring new nonlinear optical phenomena \cite{Gu2020, Mukamel2020}. Furthermore, collective coupling of $N$ molecules can be used to enhance the effective coupling strength $g$ without physical changes in the cavity design \cite{Schutz2020}. This would allow, for example, to tune in an $(N+1)st$ molecule to a cavity that is resonantly coupled to $N$ molecules. These advances will allow the engineering of new hybrid states of light and matter with fascinating properties \cite{carusotto2013quantum,haakh2016polaritonic}. \\

\textbf{Acknowledgements.} We acknowledge fruitful discussions with Claudiu Genes and thank Maksim Schwab and Fabian Greiner for technical assistance. J.N. would like to acknowledge the numerous challenges encountered during this work, which truly tested his perseverance and determination. In addition, Tobias Utikal’s help with the laser and cryogenic systems is acknowledged. This work was financed by the Max Planck Society and the German Research Foundation, Project-ID 429529648TRR 306 QuCoLiMa (Quantum Cooperativity of Light and Matter).\\ 

\textbf{Author contributions.} J.N. performed all experiments and analyzed the data reported in this paper. A.P. worked on the early theoretical and experimental developments of the project. J.R. prepared the micromirrors. S.G. co-supervised the experiment in its initial phase and contributed to its development. V.S. and S.G. conceived the original experiments. V.S. supervised the project. V.S. and J.N. wrote the paper. All authors commented on the manuscript.

\clearpage 

\bibliography{Nobakht_31122024} 

\bibliographystyle{sciencemag}

\end{document}